\documentclass[sigplan,nonacm]{acmart}
\usepackage{xspace}
\usepackage{graphicx}
\usepackage{booktabs}
\newcommand{\sysname}{\textsc{Crossflow}\xspace}
\newcommand{\sysstatic}{\textsc{Static}\xspace}

\begin{document}
\title{\sysname: Prefill--Decode Elasticity for Agentic LLM~Serving}

\author{Yi Xu, Ehsan K. Ardestani, Wenyin Fu, Martin Schatz,
  Krishna Malladi, Zhan Shu, Adnan Aziz, Shobhit Kanaujia,
  Ajit Mathews, Chunqiang Tang\\[0.5em]
  \normalfont\large\textit{Meta Platforms}}
\renewcommand{\shortauthors}{Xu et al.}

\begin{abstract}
As serving capacity demand surpasses that of training, serving efficiency
becomes increasingly important.  Prefill--decode (P/D) disaggregation improves
serving efficiency through specialization and isolation of the two phases.
These benefits rest on a static partitioning.  Phase demand,
however, is not static.  We observe that in a large LLM fleet the ratio of uncached
input to output tokens has peak-to-mean ratios up to 4.7$\times$ at minute
timescales, and that in a public agentic trace the hourly ratio spans a median
24.5$\times$ within a single day, while reassigning a replica takes tens of
minutes.  Agentic traffic sharpens the mismatch. Sizing each pool at its ninety-fifth percentile
leaves up to 17\% of cluster capacity unused; sizing below it converts the same
imbalance into queueing and unrealized throughput. 
We present \sysname, which makes this boundary elastic without changing node
roles.  Each decode node publishes a short-lived, revocable lease that bounds
local-prefill compute, KV capacity, transfer work, and projected output. Across public and internal traces, \sysname improves token
throughput by 16.2--17.4\% on geometric mean over static P/D, and by up to
43.4\% at high load, while reducing mean TTFT at every evaluated point.
\end{abstract}

\maketitle
 
\section{Introduction}

Inference serving is becoming a dominant driver of AI infrastructure
capacity, with the inference vs training crossover expected past
2027~\cite{jll2026outlook, abi2026inference}.  Agentic workloads amplify this:
per-request token consumption is 4--15$\times$ that of
chat~\cite{signal65tokenomics}, per-workflow inference computational cost is forecast to rise more than
5$\times$ by 2028~\cite{gartner2026inference}, and aggregate token volume
24$\times$ by 2030~\cite{goldman2026agents}.  At this scale, serving efficiency
is a first-order determinant of infrastructure requirements.

Serving stacks pursue this efficiency along several axes, including continuous
batching~\cite{orca,vllm}, quantization~\cite{gptq,awq}, and prefix
reuse~\cite{promptcache,sglang}.  Prefill-decode (P/D) disaggregation is
distinctive among them in focusing on the structure of the computation rather
than the model or the request
stream~\cite{splitwise,distserve}.  \emph{Prefill} processes
an input prompt and is typically compute intensive, while \emph{decode}
generates one token per active sequence in each iteration and is often
constrained by memory bandwidth and KV-cache capacity, so colocating them
forces one replica configuration to serve two different bottlenecks.  P/D
disaggregation instead assigns the phases to separate pools and transfers a
request's KV cache after prefill~\cite{distserve,splitwise,mooncake}, raising
fleet efficiency in two ways: \emph{phase specialization} lets each pool tailor
its serving configuration, and \emph{performance isolation} keeps long-running
prefill kernels from perturbing decode iterations.

\begin{figure}
    \centering
    \includegraphics[width=\linewidth]{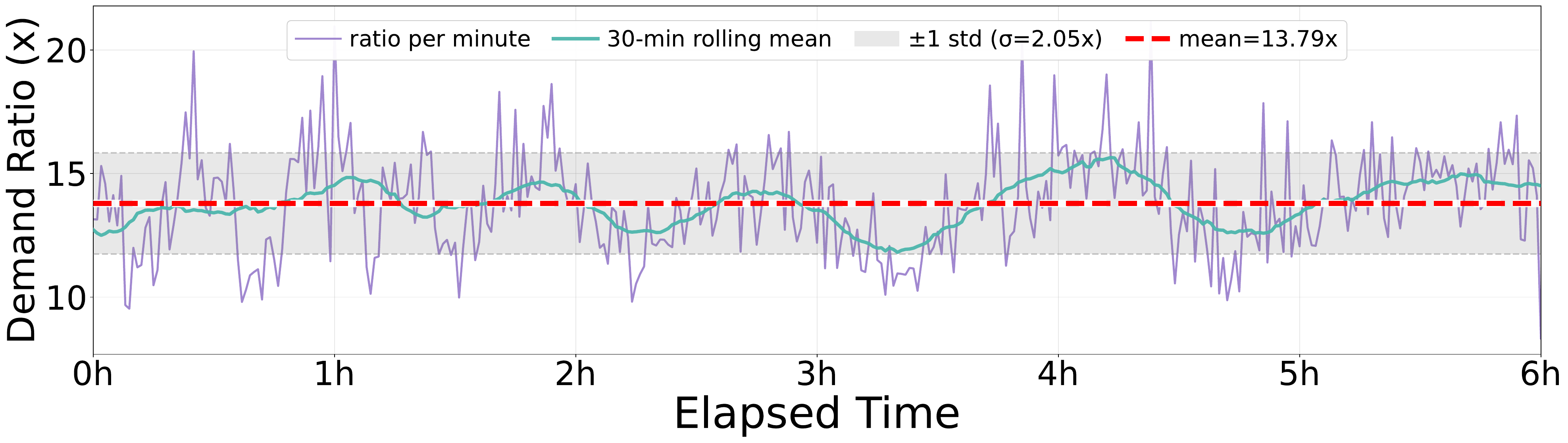}
    \caption{Demand ratio over a six-hour period for a chat use case. The rate of fine-grained change in prefill vs decode demand does not match the static partitioning nature of P/D disaggregation.}
    \label{fig:demand_ratio_time_series}
\end{figure}

Both benefits, however, are obtained through a \textbf{static partitioning}.
An operator can change the number and configuration of prefill and decode
replicas, but model initialization, communication-group setup, cache movement,
and warmup make reassignment take tens of minutes, during which the GPUs do not
serve requests.  Phase demand moves far faster.  Sampling the prefill-to-decode
load ratio for a chat use case every tens of seconds,
Figure~\ref{fig:demand_ratio_time_series} shows uncached input to output tokens
varying with coefficients of variation between 0.14 and 0.34 and peak-to-mean
ratios up to 4.7$\times$.
Nor is the problem confined to short bursts: across 134 days of a public
agentic trace, the hourly ratio spans a median 24.5$\times$ within a single
day, with only a weak time-of-day component, so even hour-scale repartitioning
would have to be planned against a ratio it cannot predict.
Agentic traffic sharpens the mismatch: turns arrive in clusters, tool
completions release dependent requests together, prefix reuse changes prefill
work, and variable output lengths change decode occupancy.  An operator must
therefore size each pool for a high percentile of its demand, stranding
cluster capacity, or size below it and let one pool queue while the other holds
idle headroom.

Our insight is that P/D node roles can remain fixed while the
\textbf{boundary between roles becomes elastic}.  We present \sysname, a P/D serving
system in which decode nodes remain decode-first but temporarily lend bounded
compute when prefill congestion and decode SLO slack coincide.  The pools keep
their phase-specific configurations, only selected prefills execute
opportunistically on a decode node, and lending falls to zero as soon as decode
needs the capacity.  This also permits deliberate asymmetric provisioning: the
prefill pool can target expected long-term demand while the decode pool retains
a conservative margin.

Lending within bounds is not the same as routing work to a lightly loaded node.
Our controlled testbed measurements show that decode throughput, compute
activity, and memory traffic change at different rates as batch occupancy and
context length vary.  Under identical aggregate decode load, alternative
placements change SLO-safe prefill throughput by 11--34\%, and the preferred
placement changes with the prefill shape.  Aggregate batch size or GPU
utilization therefore cannot identify SLO-safe capacity: it depends jointly on the
candidate prefill's work and KV footprint and on the resident decode
composition.

Prior work supplies components of this design space~\cite{splitwise, sarathi, muxwise,nexus,duetserve,taichi,ppd}. What remains unresolved is
a closed-loop capacity policy between cluster-wide admission and rapidly
changing node-local feasibility.  \sysname supplies it: each decode node
publishes a short-lived, multi-resource lease, the cluster scheduler
atomically reserves it, and a locally revalidated execution backend advances
admitted prefill work in bounded units alongside decode.

We implement \sysname in SGLang~\cite{sglang} and evaluate it on GPT-OSS-120B and GLM-5.2 on
NVIDIA GB300 servers using public~\cite{tracelab} and internal traces. Across both models and both traces, \sysname improves token
throughput by 16.2--17.4\% on geometric mean over static P/D, and by up to
43.4\% at high load, while reducing mean TTFT at every evaluated point.  The
improvement carries a bounded decode-latency computational overhead, which we quantify in
Section~\ref{sec:eval-load}.  This paper makes the following contributions:
\begin{itemize}
    \item We use large-scale fleet measurements and controlled testbed
    experiments to show that short-lived phase imbalance strands a substantial
    fraction of a static P/D configuration, and that SLO-safe lending capacity is
    nonlinear and request-shape dependent.

    \item We introduce an elastic capacity boundary for P/D disaggregation that
    preserves each node's role, phase-specific configuration, and decode-first
    execution.

    \item We design a hierarchical controller that couples computational
    lending budgets published by decode nodes with two-sided shape- and cache-aware cluster
    admission, atomic resource reservation, and backend-independent local
    revalidation.

    \item We show, through timestamp-preserving replay of both traces, that
    elastic lending improves throughput and hardware utilization across load
    levels and static P/D partitions without changing the cluster's topology.
\end{itemize}

\section{Background}
\label{sec:background}

\subsection{Prefill--Decode Serving}

For each request, prefill processes the input tokens and materializes their KV
cache before decode generates output tokens autoregressively.  Because decode
repeatedly reads model weights and an expanding KV cache, its performance is
sensitive not only to memory bandwidth but to batch occupancy, sequence length,
KV footprint, and communication latency.  This composition, rather than any
single resource, later determines how much prefill a decode node can absorb.

The phases expose different user-visible objectives.  Time to first token
(TTFT) includes admission delay, prefill queueing and execution, and any KV
transfer to a decode node.  Inter-token latency (ITL) measures the delay between
consecutively generated tokens.  A serving system must reduce prefill delay
without making active generations unresponsive.

Prefix caching separates the prompt visible to the application from the work
presented to the GPU: a long multi-turn prompt whose prefix is cached may need
only a small incremental prefill, while a shorter cold prompt may need more.
Scheduling and routing must therefore consider the tokens not already cached
and the location of reusable KV state, not prompt length alone.

\subsection{Why Disaggregate Prefill and Decode?}

P/D disaggregation assigns dedicated model replicas to each
phase~\cite{distserve,splitwise,mooncake}, yielding the two benefits noted
earlier.  Specialization is possible because separate pools may differ in
parallelism, batching policy, replica count, and even hardware type:
a configuration tuned for large prefill matrix operations is not the one that
best balances token latency, concurrency, and KV capacity for decode.
Isolation follows because dedicated decode replicas do not ordinarily execute
prefill kernels, so their token iterations avoid interference from
variable-length prompts, improving ITL predictability and tail latency.  A
decode node is therefore not a slower prefill node; it is tuned for a
different job.

\subsection{The Static Capacity Boundary}

Conventional P/D systems assign each replica to one pool, so the resulting P/D
ratio determines how much cluster capacity each phase can have.  Because expanding a pool requires
reassigning a whole replica, each pool is instead provisioned conservatively to
absorb demand variation while maintaining acceptable TTFT and ITL.

The boundary also makes routing stateful.  A prefill must deliver its KV cache
to the selected decode node, while later conversation turns benefit from cache
affinity.  Redirecting work can reduce queueing but may require cache loading,
KV transfer, or remote execution.  A useful adaptation mechanism must
therefore consider both available compute and request-specific state.

\section{Characterizing the Lending Opportunity}
\label{sec:opportunity}

Before designing such a mechanism, we ask how much capacity a fixed split
strands, and where that capacity sits.  We use normalized, aggregated
measurements from a large LLM fleet, together with a public agentic trace that
extends the analysis to longer windows.

\subsection{Static Provisioning Strands Capacity}

\paragraph{Demand ratio.}
As a proxy for the instantaneous balance between the phases we use the ratio of
uncached input tokens to output tokens,
$R(t) = \mathrm{Tok}_{\mathrm{in,uncached}}(t) / \mathrm{Tok}_{\mathrm{out}}(t)$;
counting only uncached input avoids treating a reusable prefix as new prefill
work.  This ratio characterizes relative demand, not the P/D resource ratio
itself---prefill processes tokens one to two orders of magnitude more
efficiently than decode, so $R(t)=1$ does not imply equal resource
requirements---and what matters is how far it moves from the value used to
provision a static configuration.

\begin{figure}
    \centering
    \includegraphics[width=\linewidth]{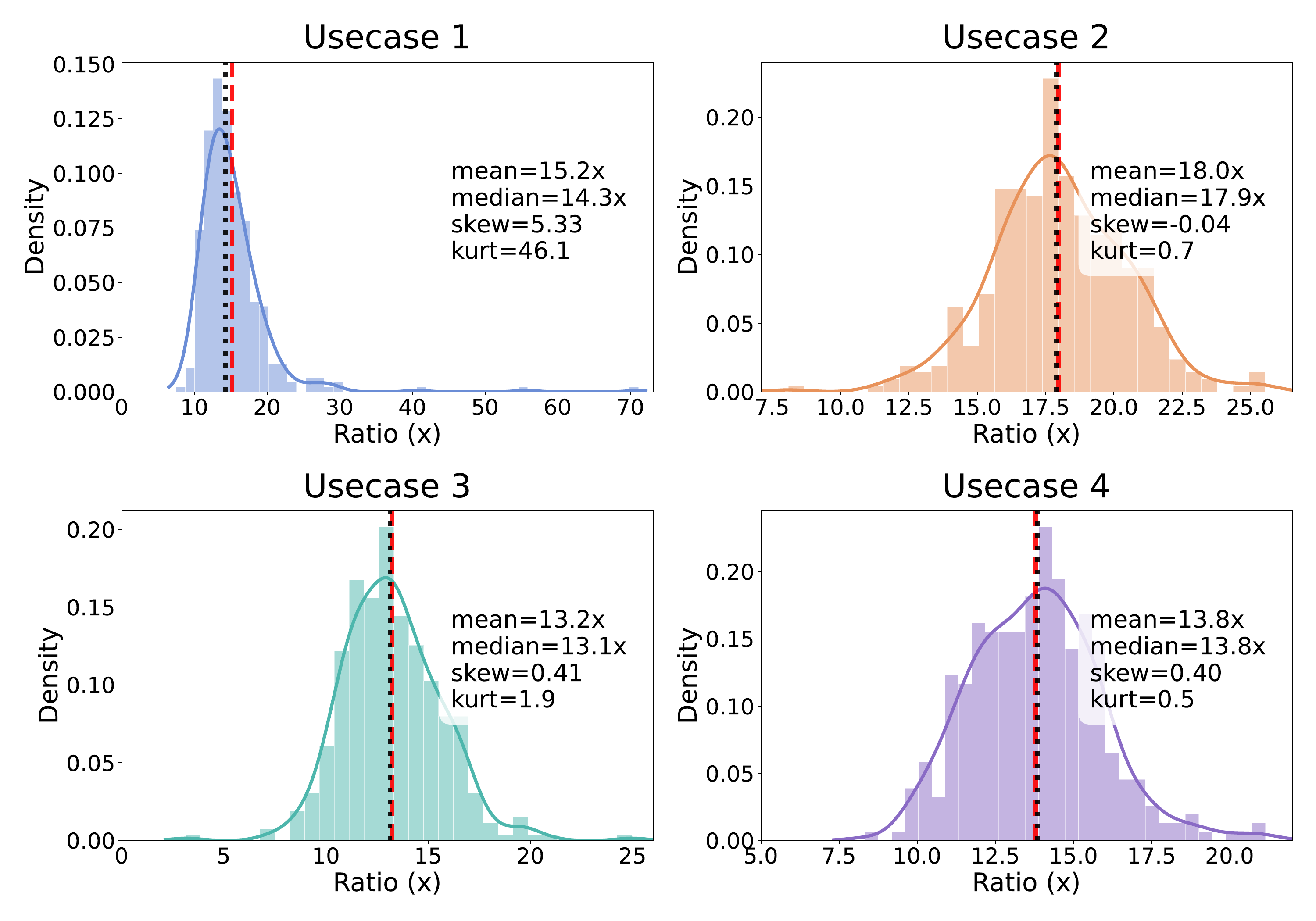}
    \caption{Per-minute demand-ratio distributions over 24 hours for three
    agentic use cases (1 - 3) and one chat use case (4).  Red dashed lines show
    means, and black dotted lines show medians.}
    \label{fig:demand_ratio_histogram}
\end{figure}

\paragraph{Minute-scale shifts.}
Figure~\ref{fig:demand_ratio_histogram} shows that the demand ratio changes
substantially at minute granularity across four use cases. Across the observed windows, demand shifts by approximately 20\%
in either direction around its central value on average, and variation
increases at shorter timescales, with one use case exhibiting a peak-to-mean
ratio of 4.7$\times$.  Even the least variable use cases, with coefficients of
variation between 0.14 and 0.18, exhibit a 1.5$\times$ to 1.75$\times$ spread between their
fifth and ninety-fifth percentiles, and the most variable use case reaches 2$\times$.
These extreme tails, together with non-trivial variability even in the
steadiest cases, demonstrate that no single static P/D ratio can efficiently
serve inference traffic with this variability.

\paragraph{Hour-scale shifts.}
Minute-scale variation motivates a mechanism that reacts within a scheduling
interval, but the mismatch persists at the hour scale, where repartitioning is
at least conceivable.  Across 2,959 hours of the public TraceLab
trace~\cite{tracelab}, the hourly uncached-input-to-output ratio swings widely
within a single day, even after each hour is normalized against its own day's
mean to remove inter-month drift.  Over 134 complete days, the median
within-day coefficient of variation is 0.89 (p90 1.58), the median within-day
max/min range is 24.5$\times$ (p90 near 100$\times$), and the typical day
contains an hour deviating 3.9$\times$ from its daily mean.  The systematic
time-of-day component is small by comparison---the median normalized hour lies
between 0.71 and 1.34---so the swing reflects bursty workload mix rather than a
predictable diurnal curve that provisioning could anticipate.

\paragraph{Opportunity size.}
$f(t) = \rho R(t) / (1 + \rho R(t))$ is the share of serving capacity prefill should hold, given $R(t)$ and $\rho$. While $R(t)$ represents workload prefill-to-decode demand ratio, $\rho$ represents the hardware decode-to-prefill
median throughput ratio.  With total demand fixed so that only the mix varies, a
configuration provisioned at the ninety-fifth percentile of each phase holds
$f_{95}$ prefill and $1 - f_{5}$ decode capacity, stranding exactly
$f_{95} - f_{5}$.  This does not average out: whenever prefill runs below its
provisioned share, decode exceeds its own by the same amount.  The four use
cases strand 17.5\%, 10.8\%, 13.8\%, and 12.0\% of a balanced configuration, and
the estimate is insensitive to the design point---at a 1:3 prefill-to-decode
split the range becomes 8\% to 14\%.  Provisioning for the observed peak rather
than the ninety-fifth percentile strands 40\% in the most variable use case.
The public trace agrees at hour scale: a split sized to its corpus mean of
roughly 12.5 uncached input tokens per output token under-provisions prefill by
4--5$\times$ in a typical day's peak hour while stranding that capacity in its
troughs.  The fleet figures are lower bounds on total stranded capacity, because they
attribute nothing to variation in aggregate load, and they bound the capacity
that any elastic mechanism can recover.

\paragraph{From stranded capacity to lending.}
When $R(t)$ rises prefill is short and decode
holds the surplus, and when it falls the surplus sits with prefill.  Because
both pools run the same model, prefill work can execute on a decode node,
making the decode-side surplus during prefill-heavy intervals the recoverable
share---and recovering it needs no replica movement, only that a decode node
serve bounded prefill work while its own SLO headroom lasts.  That matters even
at hour scale, where an operator could repartition instead.  On our testbed a
single worker takes 4:02 (GPT-OSS-120B) or 7:49 (GLM-5.2) from launch to ready,
a full 2P4D pool 8:35 or 14:59, and the cluster cannot serve until 11:02 or
17:31.  Loading weights is the small part, 19.2\,s and 127.5\,s; runtime and
cache initialization, CUDA-graph capture---7.8\,s and 171.6\,s for decode
alone---warmup, and router setup dominate.  These are end-to-end times for a
harness that also brings workers up serially, not an optimized floor, but they
place a repartition in the ten-to-twenty-minute range against a ratio that
moves within one.  The new split must also be chosen before the coming hours'
mix is known, so lending needs neither an idle period nor a forecast.

\paragraph{Asymmetric provisioning.} Lending admits a simpler provisioning rule: size prefill near its median demand,
size decode at the ninety-fifth percentile or higher, and lend decode slack to
prefill whenever prefill runs above its median.  The decode margin structurally
covers the prefill deficit, because a node's slack $f(t) - f_{5}$ is at least
the prefill shortfall $f(t) - f_{50}$ at every instant, and this asymmetric
sizing reduces stranded capacity from 11--18\% to 6.0--7.4\%.  The design
question therefore becomes instantaneous rather than long-run: how much prefill
work can a decode node absorb right now without violating its ITL SLO?  The
natural answer, lend in proportion to spare utilization, turns out to be wrong.

\begin{figure}[t]
    \centering
    \includegraphics[width=\columnwidth]{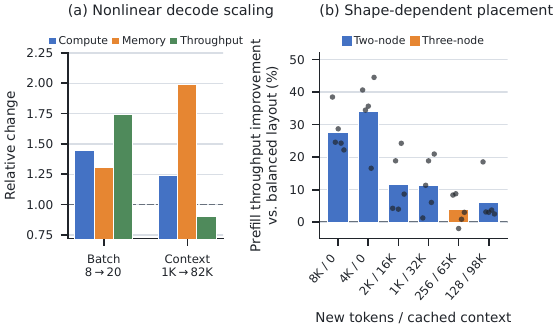}
    \caption{SLO-safe lending capacity is nonlinear and shape and model dependent
    (GPT-OSS-120B, 120 ms p99 ITL SLO).  (a)  compute, memory, and decode throughput scales with different factors.
    (b) At a fixed aggregate decode load the best
    SLO-safe placement changes with prefill shape.  Bars show prefill
    throughput relative to balanced placement; dots show five paired trials.}
    \label{fig:workload-shape-motivation}
\end{figure}

\subsection{Lending Capacity Is Shape Dependent}

A decode node with low arithmetic utilization does not expose a fixed amount
of prefill capacity.  The interference produced by local prefill depends
instead on the \emph{request shape} on both sides: for the resident decode
work, batch occupancy, context lengths, remaining output work, and KV
footprint; for the prefill work admitted beside it, the prompt tokens that
require new computation and the cached context that must remain accessible.

Figure~\ref{fig:workload-shape-motivation}(a) illustrates the nonlinearity.  A
2.5$\times$ increase in decode batch size raises compute activity by 1.44$\times$, memory
traffic by 1.30$\times$, and decode throughput by 1.74$\times$, whereas at batch 8 increasing
context from 1K to 82K nearly doubles memory traffic, raises compute activity
by only 1.24$\times$, and reduces decode throughput by approximately 10\%.  Neither
batch size nor one resource counter provides a linear measure of spare service
capacity.

Placement matters as well.  Every configuration in
Figure~\ref{fig:workload-shape-motivation}(b) carries the same aggregate decode
load; changing only its distribution across four nodes improves SLO-safe
prefill throughput by 11.3--34.0\% for the four saturated shorter-context
shapes.  Longer contexts prefer different layouts: 65K tokens favors a
three-node decode layout, while 98K tokens favors a concentrated two-node
layout.  In a separate controlled matrix, an assignment selected through
aggregate compute and memory-bandwidth complementarity was 4.24\% worse than
the empirically preferred assignment.  The preferred grouping must therefore
come from measured interference rather than a fixed resource label.

These observations motivate two levels of control.  The cluster scheduler must
select a request and destination using the candidate request shape and the
projected decode composition, while the decode node must retain control over
a short-lived, measured SLO envelope because local conditions change faster
than cluster-level routing.  Section~\ref{sec:design} shows how \sysname
couples these two levels through a single short-lived capacity lease.

\section{\sysname Design}
\label{sec:design}

\sysname makes the capacity boundary of an existing P/D configuration elastic.
Prefill and decode replicas retain their roles, model-parallel
topologies, and default request paths.  When the prefill pool becomes
congested, a decode node may temporarily accept selected prefill work, control
how quickly that work advances, and stop it when foreground decode load
changes.  A request uses the spilled path only when the prefill pool needs
relief, execution on a decode node is expected to finish earlier than the
default path, and that node signals sufficient SLO-safe capacity.
These three conditions, \emph{demand}, \emph{benefit}, and \emph{SLO
constraints},
structure the rest of this section.  If any condition fails, the request
follows the default disaggregated path.

\begin{figure}[t]
    \centering
    \includegraphics[width=\columnwidth]{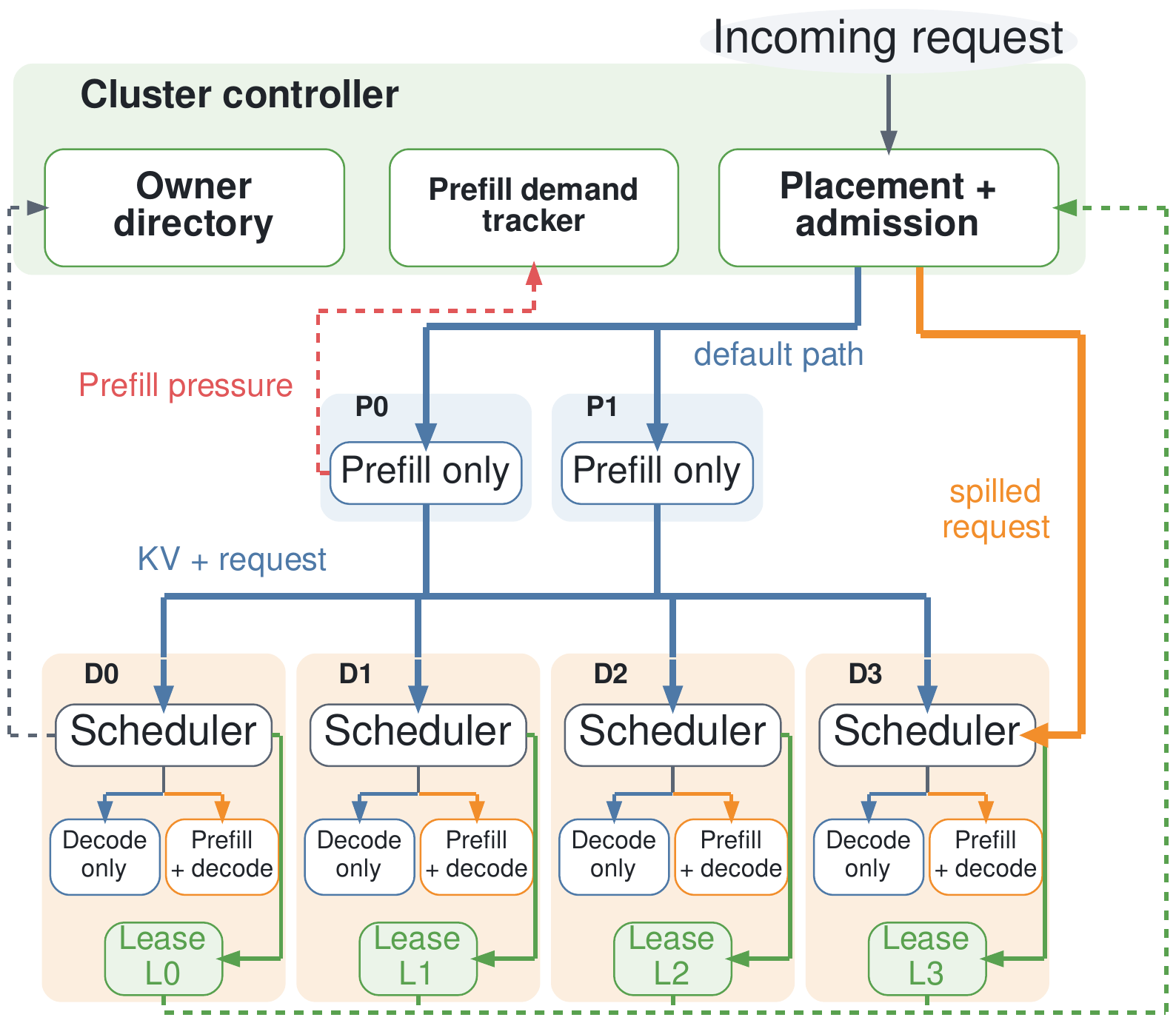}
    \caption{System architecture for a 2P4D configuration.  Solid lines show the
    default and spilled request paths.  Dashed lines carry prefill-pressure
    reports and decode-capacity leases; successful completions update the owner
    directory.}
    \label{fig:crossflow-system-architecture}
\end{figure}

During normal operation, the cluster controller sends prompt work through the
prefill tier before generation continues on the decode tier.  Under congestion,
the prefill tier reports queue pressure while each decode node signals a
short-lived, multi-resource lease $L_d(t)$ covering compute, KV capacity,
projected output work, transfer capacity, and spill slots.  The controller
redirects a request only after reserving sufficient leased capacity, and the
selected prefill then executes alongside decode work on that node.  Per-node
feasibility does not imply cluster-wide feasibility, however: if every decode
node lends at once, the reserve needed by requests about to arrive from the
prefill tier disappears, so \sysname pairs these leases with a cluster-wide
borrowing limit.  Figure~\ref{fig:crossflow-system-architecture} summarizes
these request paths and control signals.

\subsection{Detecting Prefill Pressure}
\label{sec:design-prefill-pressure}

Polling only the number of waiting requests is too reactive for short bursts:
the request that creates a queue may already have been routed before the next
poll reports congestion, and request counts ignore the variation in incremental
prompt work caused by prefix caching.  \sysname therefore combines observed
and predicted pressure.  The cluster scheduler polls the prefill workers and
maintains a virtual queue between polls, charging each arrival only the prompt
tokens that require new computation so a long cached conversation does not
appear more computationally expensive than a shorter cold prompt.  With
$W_P(t)$ the aggregate outstanding prompt work across the prefill tier and
$\widehat{\mu}_P(t)$ its aggregate recent processing rate, the estimated wait is
$\widehat{Q}_P(t)=W_P(t)/\widehat{\mu}_P(t)$, and completed requests update
$\widehat{\mu}_P(t)$.  Charging a request on arrival rather than at the next
poll, and releasing it once a decode node reserves it, lets the request that
forms a burst trigger lending instead of waiting for a later sample.  The
pressure signal becomes true when a worker reports queued prefill work or
$\widehat{Q}_P(t)$ exceeds an activation threshold, and remains true until
$\widehat{Q}_P(t)$ holds below a lower threshold for a bounded number of
updates; direct queue observations correct drift.  If a prefill worker cannot
report valid state, \sysname
disables new lending rather than interpreting missing telemetry as spare
capacity.

\subsection{Deriving an SLO-Safe Computational Lending Budget}

As Section~\ref{sec:opportunity} showed, a decode node cannot derive an
SLO-safe computational lending budget from arithmetic utilization alone.
Spilled prefill also consumes
memory bandwidth and communication resources, while the redirected request
adds KV state and future decode work.  The same prefill can therefore be
SLO-safe beside a lightly occupied decode batch but SLO-unsafe beside one with
longer contexts or more remaining output, and SLO-safe on a node at one instant
but not the next.

\paragraph{Multi-resource lease.}
Only a decode node sees the resources a spilled prefill would share with
decode, so each computes a short-lived lease $L_d(t)$ from its live state: a
resource vector of compute, KV, projected-output, transfer, and spill-slot
credits, with a version and expiration time.  Compute credit uses units that the execution
backend can bound, such as token chunks or layer-tokens.  Each availability
signal supersedes the previous version, so unused capacity cannot accumulate over
time.

\paragraph{Computing the local envelope.}
The node derives its lease from the running decode batch, projected remaining
output, physical KV headroom, pending transfers, active and reserved spills,
health, and execution layout.  Existing reservations reduce the signaled
credits, and missing state, an unhealthy worker, or exhaustion of any hard
resource limit produces a zero lease.

\paragraph{Online feedback.}
Completed spills update the observed local prefill rate, while intervals
without spilled work establish each decode node's tail-latency baseline.  When
prefill pressure persists, decode latency remains near that baseline, and
completions keep pace with arrivals, the controller gradually increases the
compute and spill-slot credits in subsequent leases.  Decode-latency pressure
restricts those credits, and a hard SLO violation sets the next lease to
zero.  If an expansion reduces completion efficiency, subsequent leases
restore the previous limits.  As decode pressure rises, signaled capacity
approaches zero and \sysname approaches static prefill/decode operation.

\subsection{Shape-Aware Cluster Admission}
\label{sec:admission}

A valid lease exposes capacity, but it does not identify which request should
use that capacity or whether the cluster should lend it.  Admission must
preserve cache affinity, compare the two execution paths, and protect the
aggregate decode reserve.  Shape enters twice: when a conversation first
appears the controller assigns its prefill and decode \emph{homes}, the nodes
that serve its prompt and generate its output across later turns, and under
prefill pressure it chooses which waiting prefill, if any, should execute on
its decode home.

\paragraph{Shape-aware home assignment.}
Because later turns return to the same decode home, one placement decision
fixes that node's future batch composition, which
Section~\ref{sec:opportunity} showed governs how much prefill it can absorb.
\sysname estimates a new conversation's prompt work, output work, and turn
count from the current request and completed conversations of similar prompt
length and requested output limit, falling back to the current prompt,
requested output limit, and one turn when no such observations exist;
successful completions update these estimates.

Load balance controls placement first: prefill-home selection balances
projected prompt work, and decode-home selection balances projected output work
together with projected turn count, which captures per-request scheduling work
that output-token totals omit.  Shape affinity only breaks ties among nodes
within a configured load margin, steering long-context conversations toward
nodes already serving similar contexts and short prompts away from
long-context cohorts, so decode batches stay more homogeneous without ever
overriding a material load difference.

A conversation keeps its decode home across both paths.  On successful
completion the owner directory records that home together with the longest
prefix processed so far, its \emph{coverage}.  Coverage does not imply that
the corresponding KV pages remain resident, so admission reserves enough
capacity to recompute missing state.

\paragraph{Path- and shape-aware spill selection.}
Pressure alone does not make the spilled path useful.  For request $r$ with
home decode node $d$, \sysname requires the spilled path to be strictly faster:
$T_D(r,d,t) = Q_d(r,t) + S_d(r,L_d(t))$, the predicted wait at that node plus
its service time under the current lease, must fall below
$T_P(r,t) = \widehat{Q}_P(t) + S_P(r) + C_{P\rightarrow D}(r)$, the default
path's predicted prefill wait, prefill service, and KV-transfer time.  The
decode-side service estimate covers the prompt suffix beyond the
completion-derived coverage, or the full prompt without a valid coverage
record.  Both sides use only the prompt, requested output limit, and
completed-prefix record available at admission.  Otherwise the request stays on
the default path, and \sysname does not move an established conversation
between decode nodes to create a spill opportunity.

Arrival order is not the best spill order, because removing a request reshapes
the forming prefill batch while inserting it changes the target node's
coexecution efficiency and future decode load.  \sysname therefore collects a
bounded set of waiting candidates over a short interval and ranks each by

\[
  \mathsf{score}(r,d)
  = T_P(r,t)-T_D(r,d,t) + B_P(r,t)-I_D(r,d,t),
\]

The term $T_P(r,t)-T_D(r,d,t)$ measures the selected request's predicted
completion-time benefit.  The remaining terms capture its effect on other
requests: $B_P(r,t)$ estimates the benefit to requests left in the forming
prefill batch, while $I_D(r,d,t)$ estimates interference imposed on active
decode requests from suffix size, differences in context length and requested
output limit, active peers, and consumption of otherwise idle decode headroom.
Absent observations the scheduler favors less new prompt work and smaller
context- and output-length differences from active requests on the target node;
completed spills then refine these computational costs from measured time to
first token,
each correction bounded so sparse estimates stay near the suffix-size average.
The score orders candidates but cannot override the admission constraints.

\paragraph{Cluster-wide SLO-safe admission.}
The controller pairs each node's lease with a cluster-wide concurrency limit
$G(\widehat{Q}_P(t))$, zero below the pressure threshold and rising in stages
as predicted prefill wait grows, with higher concurrency demanding stricter
request-shape and cluster-headroom checks.  With $K_d(t)$ the spill-slot
allowance in node $d$'s current lease and $n_d(t)$ its active or reserved
spills, admission requires both $n_d(t)<K_d(t)$ locally and
$\sum_d n_d(t)<G(\widehat{Q}_P(t))$ cluster-wide.  A candidate must also target
its assigned decode home, observe prefill pressure, satisfy
$T_D(r,d,t)<T_P(r,t)$, fit the request-shape bounds, and fit every compute, KV,
projected-output, and transfer credit in the lease.  The controller takes
candidates in score order, admits the first that satisfies every condition, and
reserves its credits before considering another, so concurrent admissions
cannot overcommit the same capacity.

\subsection{Decode-Node Execution and Recovery}

The cluster controller decides whether a request may use the spilled path; the
decode-side executor determines how the reserved work shares the node with
decode.  \sysname does not require one particular coexecution mechanism.  It
requires only that the executor bound each unit of prefill work, revalidate
local state before issuing it, pause unissued work when conditions change, and
account for the resources covered by the lease; Section~\ref{sec:implementation}
describes the backends we built against this interface.

\paragraph{Spilled-request execution.}
The controller sends the complete request to its assigned home decode node,
where a radix-cache lookup determines how much of the prompt is reusable and
the executor processes the remaining tokens; a cache miss makes that work the
full prompt.  Because the executor revalidates before issuing each unit, a
reservation permits a request to wait at the decode node but does not force
execution after local conditions change.

\paragraph{Failure handling.}
\sysname is conservative under failure.  A request that cannot obtain a valid reservation stays
on the default disaggregated path.  If decode conditions become SLO-unsafe after
admission, the decode scheduler pauses unissued prefill work.  A failure before
the first token releases the reservation and retries the request through the
prefill tier.  After the first token, the request remains on its decode node
until generation terminates.  Only successful completion advances the prefix
coverage record.  This boundary prevents duplicate output and ambiguous cache
ownership.

\section{Implementation}
\label{sec:implementation}

We implement \sysname in SGLang~\cite{sglang}.  A user-space Python shim of
6.7K lines wraps the existing cluster router and implements the virtual prefill
queue, home assignment, path comparison, lease reservation, and failure
recovery.  Worker-side support adds 2,807 lines and removes 36 across 15 SGLang
files, mostly extending the disaggregated-decode manager and scheduler or
adding a lease-accounting module; smaller hooks expose load and completion
events, validate cache and resource availability, and release reservations.
The implementation requires no new model kernels.

SGLang's disaggregated decode path normally receives a request only after a
prefill worker has produced its KV state.  We extend this path with a local
prefill queue, allowing a decode worker to accept the complete request, reuse
any resident prefix, and construct a mixed prefill/decode batch.  The existing
prefill-to-decode path remains unchanged.  Because logical cache ownership can
outlive physical residency, the worker checks its radix cache and physical KV
capacity before dispatch.  For tensor-parallel replicas, all ranks validate
the same lease before issuing the request.

We implement the executor interface twice: with SGLang's native chunked
mixed-batch path, and with PDMUX~\cite{pdmux}, which spatially partitions
decode-node resources between prefill and decode.  A MuxWise-style
executor~\cite{muxwise} would satisfy the same interface.  The two backends
reach similar end-to-end performance, and all results in
Section~\ref{sec:evaluation} use the native chunked mixed-batch path.

\section{Evaluation}
\label{sec:evaluation}

Our evaluation asks how \sysname affects throughput, TTFT, and decode ITL;
whether its improvements hold across models, traces, and P/D partitions and against
prior policies; how much static hardware is needed to match it; and which
control mechanisms matter.

\subsection{Experimental Setup}

\paragraph{Models and cluster.}
We evaluate GPT-OSS-120B (117B total and 5.1B active parameters per token) and
GLM-5.2 (753B total and 40B active) on NVIDIA GB300~\cite{nvidia_gb300} servers, each with four
288-GB GPUs and one TP4 worker.  Unless otherwise stated, experiments use six
servers in a 2P4D configuration.  Unless varied explicitly, comparisons use the
same checkpoint, precision, server count, P/D split, KV-cache capacity, and
distributed L3 cache.  Sections~\ref{sec:eval-partitions}
and~\ref{sec:eval-static-scaling} vary the split and server count.

Workers use a June 18, 2026 SGLang snapshot (kernel 0.4.2.post2), a 128K-token
context with 512 tokens reserved, at most 256 running requests, a 0.35 static
GPU-memory fraction, and a 100-GB write-through Mooncake cache.  NVLink handles
TP4 communication, NIXL~\cite{nixl} transfers P-to-D KV state, and RDMA accesses L3.
To make the comparison isolate the scheduling policy, matched runs use a common
backend configuration.  We leave optional backend optimizations disabled in
these experiments (e.g., custom all-reduce).

\begin{figure*}[t]
    \centering
    \includegraphics[width=\textwidth]{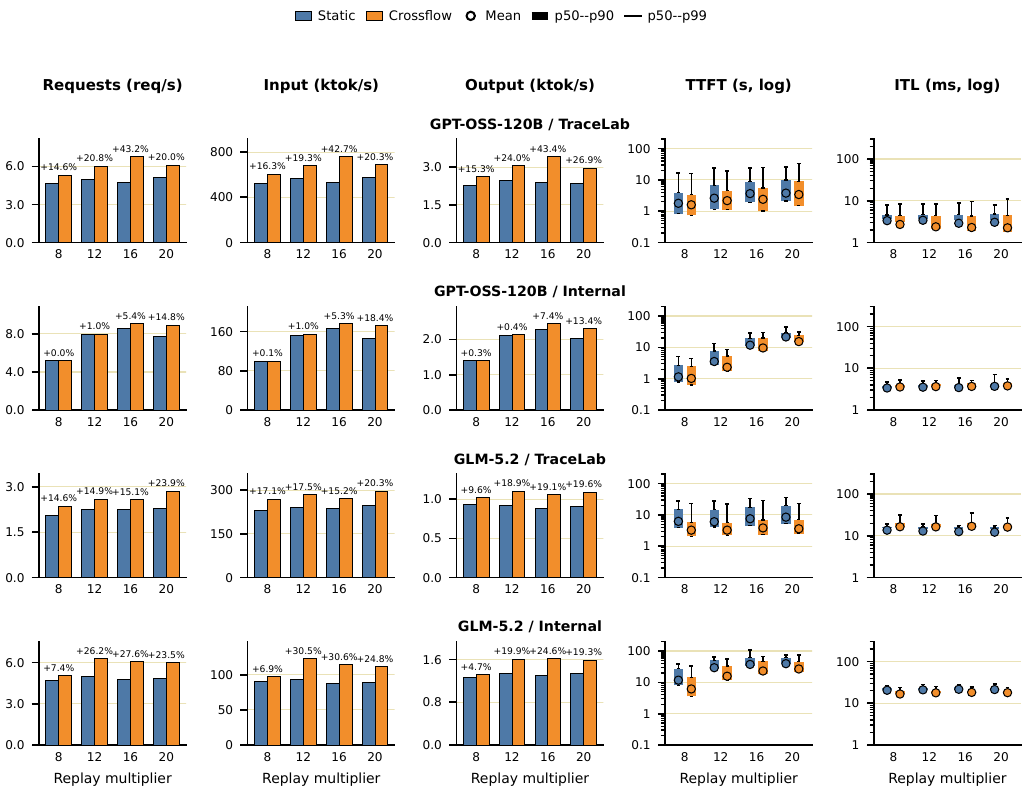}
    \caption{End-to-end sustained replay with matched 2P4D configurations.
    Rows identify the model and trace; columns report request, input-token, and
    output-token throughput, TTFT, and request-mean ITL.  Latency glyphs show
    the mean (circle), p50--p90 (thick), and p50--p99 (thin).
    Multipliers scale each trace's native arrival rate and are not comparable
    across traces.}
    \label{fig:eval-load-sweep}
\end{figure*}

\paragraph{Timestamped trace replay.}
We use one-hour segments from the public TraceLab coding-agent trace and an
internal trace, preserving timestamps, conversation order, prefix
relationships, and token work.  A load factor $\alpha$ divides every timestamp
offset from the first arrival by $\alpha$, increasing rate without changing the
trace's temporal shape.

\paragraph{Baselines.}
All baselines use the same SGLang substrate, cache capacity, and server count.
\textbf{\sysstatic} uses the same fixed P/D split as \sysname and sends every
prefill to a P node.
\textbf{PPD}~\cite{ppd} dynamically routes later-turn append-prefills to D
nodes holding their KV state, making it the closest baseline.  Its routing table
requires offline profiling for each workload shape and load level; following
its methodology, we calibrate it for every workload and replay rate.
\textbf{SMetric}~\cite{smetric}
load-balances each session's first request and routes later turns by cache
affinity, spreading sessions while preserving later-turn reuse.
\textbf{CacheWise}~\cite{cachewise} prioritizes requests with reusable prefixes
and evicts KV entries based on predicted reuse, reducing recomputation under
cache pressure.
\subsection{End-to-End Benefit over \sysstatic}
\label{sec:eval-load}

We first compare \sysname with \sysstatic across sustained timestamped replay
sweeps, covering both models and both traces.  We report both request and token
throughput; the latter weights
heterogeneous requests by their input and output volume instead of treating
every request equally.

Figure~\ref{fig:eval-load-sweep} reports the sustained-load sweeps.  Across the
16 matched load points, the geometric-mean improvements are 16.6\% for
requests, 17.4\% for input tokens, and 16.2\% for output tokens.  The largest
input and output improvements are 42.7\% and 43.4\%, respectively.  These span
both token directions: \sysname improves input and output throughput together,
rather than trading one for the other.
The sweep creates progressively deeper prefill pressure: for GPT-OSS-120B on
TraceLab, \sysname's median virtual prefill-queue signal rises by
3.4$\times$ from 8$\times$ to 20$\times$, with the same upward trend appearing
in the other model and trace configurations.

Mean TTFT improves at every point: it falls by 10.0--34.4\% for GPT-OSS-120B on
TraceLab, by 10.7--34.4\% on the internal trace, by 46.1--57.4\% for GLM-5.2 on
TraceLab, and by 32.2--47.7\% for GLM-5.2 on the internal trace.  TTFT
improves consistently because routing and lending reduce prefill queueing.

ITL depends on whether queue relief outweighs interference between local
prefill and active decode batches.  For GPT-OSS-120B on TraceLab, useful spills
relieve pressure while leases preserve enough decode headroom, reducing mean
ITL by 18.9--30.3\%.  On the internal trace, mean ITL instead rises by
0.09--0.23 ms, or 2.5--6.7\%.

For GLM-5.2, \sysname raises mean ITL by 20.9--34.3\% on TraceLab because local
prefill interferes more with active decode.  On the internal trace, mean ITL
drops by 15.2--18.7\%.  Spills happen less frequently under this setup,
suggesting that this improvement comes mainly from routing and queueing rather
than from spilled prefill.

\subsection{Robustness to the Static P/D Partition}
\label{sec:eval-partitions}

We next test whether lending depends on one favorable static partition.  We
compare 1P5D, 2P4D, 3P3D, and 4P2D, spanning prefill- to decode-constrained
allocations, against \sysstatic at the same split.

\begin{figure*}[t]
    \centering
    \includegraphics[width=\textwidth]{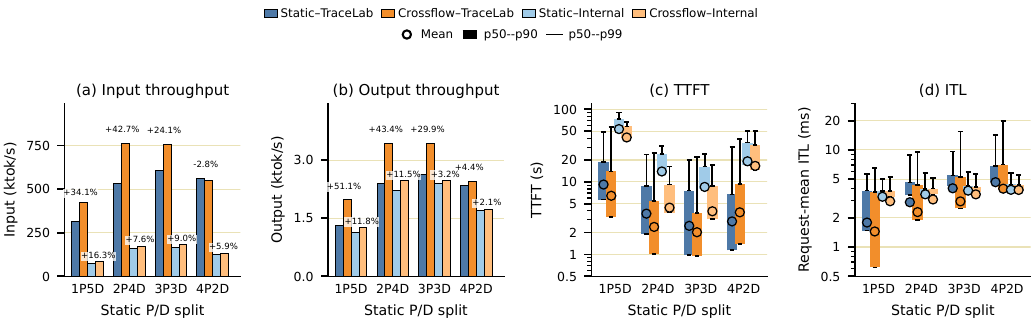}
    \caption{Robustness to the static P/D split for GPT-OSS-120B at
    16$\times$ on six GB300 servers.}
    \label{fig:eval-partition-robustness}
\end{figure*}

On TraceLab, Figure~\ref{fig:eval-partition-robustness} shows that \sysname
improves input-token throughput by 24.1--42.7\% and output-token throughput by
29.9--51.1\% at 1P5D through 3P3D.  Its best result, at 2P4D, is 25.1\% and
30.0\% above the best \sysstatic input- and output-token throughput at any
split.  Mean TTFT falls by 18.3--34.4\% and mean ITL by 19.1--26.7\% over these
three splits.  At 4P2D, however, input throughput falls by 2.8\% and mean TTFT
rises by 33.3\%, although output throughput remains 4.4\% higher.  With only two
decode nodes, the elastic lending capacity is limited.

On the internal trace, \sysname improves input-token throughput by 5.9--16.3\%
and output-token throughput by 2.1--11.8\% at every split.  Its best result, at
3P3D, is 9.0\% and 3.2\% above the best \sysstatic input- and output-token
throughput.  Mean TTFT is 13.6--68.3\% lower and mean ITL is 0.4--11.5\% lower;
p99 TTFT is 1.4--47.9\% lower, while p99 ITL ranges from 5.0\% higher to 11.6\%
lower.

The improvements therefore hold across a broad range of partitions.
Workload balance changes over hours, days, and longer periods, so the best
static P/D split can change as well.  With \sysname, this does not necessarily
require an immediate physical repartition: for example, at 2P4D, \sysname
exceeds 3P3D \sysstatic in both token-throughput metrics and has lower mean TTFT
and ITL on both traces.  Thus, even when a workload shift would favor changing
\sysstatic from 2P4D to 3P3D, \sysname can retain the existing split and avoid
the idle period that repartitioning requires.

\begin{figure}[t]
    \centering
    \includegraphics[width=\columnwidth]{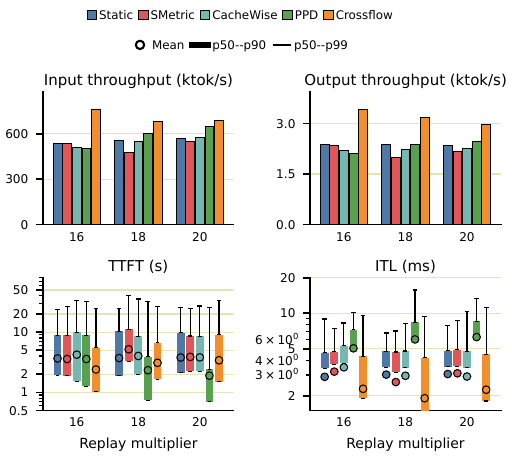}
    \caption{Comparison with state-of-the-art baselines using GPT-OSS-120B and
    a physical 2P4D configuration on TraceLab at 16$\times$, 18$\times$, and
    20$\times$ replay.  PPD denotes calibrated dynamic PPD.}
    \label{fig:eval-prior-policies}
\end{figure}
\subsection{Comparison with State-of-the-Art Baselines}

We compare \sysname with calibrated dynamic PPD, SMetric, CacheWise, and
Static under the same 2P4D allocation.  PPD is the closest prefill-offloading
baseline; SMetric and CacheWise test whether session-centric load balancing
and cache affinity alone capture the benefit of \sysname's capacity-aware
controller. Figure~\ref{fig:eval-prior-policies} shows the results.

On TraceLab, \sysname provides the highest measured token throughput at all
three rates, improving output-token throughput over the best non-\sysname
policy by 43.4\%, 32.6\%, and 20.9\%.  It also has the lowest mean ITL at all
three points: 2.29, 1.91, and 2.25 ms, compared with the best baseline values
of 2.89, 2.61, and 2.91 ms.

The PPD trace explains the remaining gap.  Its dynamic policy aggressively
offloads prefill to decode-local execution: the offload path handles
32.4\%, 76.9\%, and 90.9\% of measured requests at 16$\times$, 18$\times$,
and 20$\times$, respectively, or 26.0\%, 80.9\%, and 92.2\% of prompt tokens.
At 16$\times$, this 32.4\% offload share does not improve throughput, and only
marginally changes TTFT: PPD reaches 2.11 ktok/s output throughput and
3.57-second mean TTFT, versus 2.19--2.39 ktok/s and 3.57--4.22 seconds for the
other non-\sysname baselines.  At higher rates, PPD achieve lower TTFT than \sysname, but it comes with the tradeoff of ITL to 5.04--6.28 ms, which is 1.7--2.1$\times$ the
Static baseline and 2.2--3.1$\times$ \sysname.

\subsection{Hardware Requirements of Static P/D Provisioning}
\label{sec:eval-static-scaling}

The fixed-server-count experiments in Section~\ref{sec:eval-partitions} compare
policies under matched server counts.  We next ask how much additional
hardware \sysstatic requires to approach a smaller \sysname configuration.  We
take \sysname on six servers as the reference and evaluate \sysstatic on six,
seven, eight, and nine servers, sweeping the available P/D partitions at each
server count.  All runs use GPT-OSS-120B, the TraceLab 16$\times$ replay, and
the same sustained measurement window.

\begin{figure}[t]
    \centering
    \includegraphics[width=\columnwidth]{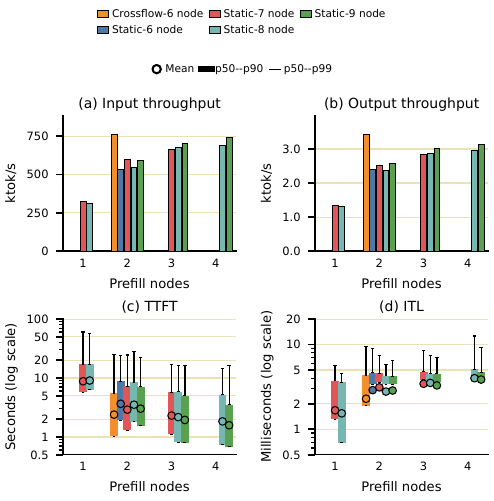}
    \caption{Hardware required for \sysstatic to approach the performance of
    six-node (2P4D) \sysname under the GPT-OSS-120B TraceLab 16$\times$ replay.}
    \label{fig:eval-static-overprovisioning}
\end{figure}

Figure~\ref{fig:eval-static-overprovisioning} shows that the matched six-node
\sysstatic 2P4D configuration reaches 534K input tokens/s, 2.39K output
tokens/s, and 4.74 requests/s, while six-node \sysname reaches 763K input
tokens/s, 3.43K output tokens/s, and 6.79 requests/s.  Static needs seven nodes
to match \sysname's TTFT: its best
seven-node split reaches 2.30/16.69 s mean/p99 TTFT, close to \sysname's
2.40/24.75 s, but still produces only 2.83K output tokens/s.  Even with nine
nodes, the best static split reaches only 742K input tokens/s and 3.15K output
tokens/s, 2.7\% and 8.3\% below \sysname while using 50\% more hardware.

We analyzed the trace.  Static provisioning hard-wires the P/D split:
decode-heavy splits keep ITL low but queue prefill, while prefill-heavy splits
lower TTFT by removing decode capacity.  In contrast, the six-node \sysname
2P4D trace sends 346 of 1,528 measured requests through the local spill path
and leaves the remaining 1,182 on the normal prefill route.  This lets \sysname
borrow decode slack for prefill without permanently dedicating more servers to
prefill.  Nine-node static still does not beat \sysname's typical ITL: all
nine-node splits have worse mean and p50 ITL, although their p99 ITL is mixed.
This is likely because \sysname admits spill work through shape and pressure
checks, while static must continuously operate at a fixed resource split.

\subsection{Ablation Study}
\label{sec:eval-ablation}

We isolate the controller mechanisms using GPT-OSS-120B, the TraceLab replay,
and a 2P4D configuration at four offered loads.  \emph{Static} disables both
shape-aware placement and spilled prefill.  \emph{Routing only} enables
shape-aware placement but retains the default disaggregated execution path.
\emph{Routing + Spill} admits resident local prefill in FIFO order without
shape-aware selection.  \emph{Routing + Shape-aware Spill} adds
shape-compatible cache-resident spilling with one fixed cluster-wide spill
slot.  \emph{\sysname} keeps shape-compatible spilling and changes
cluster-wide spill concurrency with measured prefill pressure.

\begin{figure}[t]
    \centering
    \includegraphics[width=\columnwidth]{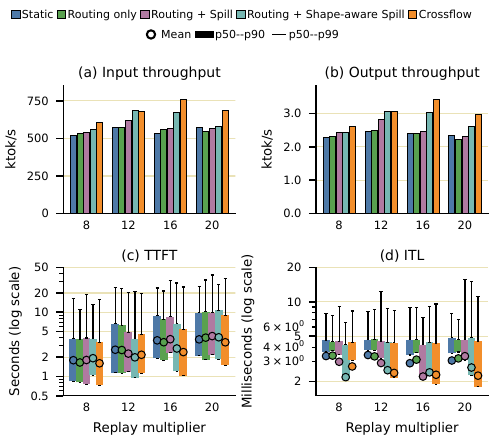}
    \caption{Controller ablation using GPT-OSS-120B, TraceLab, and a 2P4D
    configuration.}
    \label{fig:eval-ablation}
\end{figure}

Figure~\ref{fig:eval-ablation} separates placement from the ability to execute
prefill on decode nodes.  Routing alone changes input-token throughput by
$-4.3$--4.6\% and output-token throughput by $-4.6$--1.1\% relative to Static.
Routing + Spill helps at some loads but remains below \sysname by 8.6--26.0\%
in input throughput and 7.7--27.9\% in output throughput because FIFO
eligibility does not account for shape or path benefit.  The fixed one-slot Routing +
Shape-aware Spill policy is stronger, improving output-token throughput by
6.7--26.2\% over Static while reducing mean ITL by 13.2--34.7\%.

The full improvement comes from combining spill execution, shape-aware spill
selection, and a pressure-aware spill concurrency limit.  Shape-aware spill uses decode headroom more
efficiently, causing less interference with the default decode path and
preserving more usable headroom for subsequent spill admissions.  At
16$\times$ and 20$\times$, \sysname completes 346 and 240 spills, compared with
101 and 74 for Routing + Spill and 254 and 170 for Routing + Shape-aware Spill.
Relative to Static, \sysname improves input-token throughput by 16.3--42.7\%
and output-token throughput by 15.3--43.4\%, reduces mean TTFT by
10.0--34.4\%, and improves mean ITL by 18.9--30.3\%.

We also test whether the two control-plane layers are necessary
(Table~\ref{tab:eval-control-plane-ablation}).  Replacing leases with exact
synchronous D-capacity polling uses the same capacity calculation, but reduces
throughput from 6.79 to 5.68 requests/s because each admission waits on all
decode workers, adding 202\,ms of mean polling latency.  Disabling the
arrival-side virtual queue exposes a different failure mode: at 16$\times$,
periodic polling completes only 56 local spills and 5.27 requests/s, compared
with 346 spills and 6.79 requests/s for \sysname.  Thus, leases make capacity
checks lightweight enough for the admission path, while the virtual queue makes burst
demand visible before it reaches workers.

\begin{table}[t]
    \centering
    \caption{Control-plane ablations at 16$\times$ replay.}
    \label{tab:eval-control-plane-ablation}
    \scriptsize
    \setlength{\tabcolsep}{3.0pt}
    \resizebox{\columnwidth}{!}{%
    \begin{tabular}{lcccc}
        \toprule
        \textbf{Policy} & \textbf{Req/s} & \textbf{TTFT mean/p99} &
        \textbf{ITL mean/p99} & \textbf{Spills} \\
        \midrule
        \sysname & 6.79 & 2.40/24.75\,s & 2.29/9.56\,ms & 346 \\
        No virtual queue & 5.27 & 3.10/20.63\,s & 2.92/7.99\,ms & 56 \\
        Sync D polling & 5.68 & 3.11/34.30\,s & 2.49/10.88\,ms & 253 \\
        \bottomrule
    \end{tabular}%
    }
\end{table}

\section{Related Work}
\label{sec:related-work}

\paragraph{LLM serving systems.}
LLM serving systems improve efficiency through batching, memory management,
cache reuse, and model-level optimizations.  Continuous batching,
iteration-level scheduling, paging, offload, and dynamic allocation increase
throughput and effective model/KV capacity~\cite{orca,vllm,flexgen,pie_offload,deepspeed_inference,deepspeed_fastgen,vattention,flashinfer}.
Attention kernels and structured runtimes reduce operator and application
computational overheads~\cite{flashattention,flashattention2,sglang,parrot},
while quantization and adapter-serving systems reduce per-token or per-tenant
computational cost~\cite{gptq,awq,smoothquant,llmint8,punica,slora}.  \sysname is
orthogonal: it targets the capacity boundary between already-specialized
prefill and decode pools.

\paragraph{Prefill--decode disaggregation.}
P/D disaggregation separates prompt processing from token generation so each
phase can use a configuration suited to its bottleneck, improving goodput,
latency, and KV-cache management~\cite{distserve,splitwise,mooncake,memserve,greenllm}.
Other systems tune model parallelism, heterogeneous placement, energy,
autoscaling, or online reconfiguration~\cite{alpaserve,melange,dynamollm,chiron,flying_serving}.
These approaches adapt at static, replica, or engine-configuration granularity.
\sysname keeps pool assignment fixed and borrows bounded prefill execution from
decode nodes only while decode slack is available.

\paragraph{Colocated prefill and decode execution.}
Colocated systems chunk prefill work or use intra-GPU partitioning, adaptive
multiplexing, and unified aggregation to manage the prefill/decode
throughput-latency tradeoff~\cite{sarathi,muxwise,nexus,duetserve,taichi}.
Queueing, migration, and memory managers address related SLO, fairness, and
memory-pressure problems~\cite{qlm,llumnix,fastswitch,mell,jenga,kunserve}.
These works mainly decide how a node shares resources once both phases are
assigned to it; \sysname adds the cluster-level decision of when a
disaggregated cluster should borrow decode-local prefill capacity.

\paragraph{Routing, cache affinity, and agentic workloads.}
Routing must also account for sessions, cache locality, and bursty agentic
workloads~\cite{tracelab,servegen,smetric,cachewise}.  Stateful, shared-prefix,
and cache-streaming systems reduce repeated context work~\cite{promptcache,cachedattention,pensieve,relayattention,hydragen,cachegen},
while agentic-serving systems exploit program structure, tool-call idle periods,
and workflow-level state~\cite{thunderagent,mori,continuum}.  Speculative decoding
accelerates generation
orthogonally~\cite{speculative_decoding,specinfer,medusa,adaspec}.  PPD is
closest in deflecting prefill work using decode-side state~\cite{ppd}.
\sysname differs by exposing decode capacity as
short-lived multi-resource leases, reserving them atomically at the cluster
scheduler, and revalidating them locally before each bounded prefill unit.

\section{Conclusion}

P/D disaggregation improves serving efficiency through phase specialization
and performance isolation, but its static capacity boundary cannot follow
short-lived changes in prefill and decode demand.  \sysname makes this
isolation elastic.  Decode nodes remain decode-first while exposing
short-lived, revocable lending capacity.  A cluster scheduler admits and
places prefill requests against this capacity, and each decode scheduler
applies its resource reservations through a bounded, locally revalidated
execution backend.  These decisions incorporate request shape on both sides of
the spill, cache locality, and changing decode conditions instead of treating
idle compute as a scalar resource.

We implement \sysname in SGLang for GPT-OSS-120B and GLM-5.2 and evaluate it
with timestamp-preserving replays of the public TraceLab workload and an
internal trace.  Across both models and both traces, \sysname
improves token throughput by 16.2--17.4\% on geometric mean over static P/D,
and by up to 43.4\% at high load, while reducing mean TTFT at every evaluated
point.  The improvement holds across static P/D partitions, and it adds minimal
decode latency where lending is useful and more where it is not.
These results demonstrate that a P/D configuration can recover transient stranded
capacity without discarding phase specialization or decode-first isolation.

\section*{Acknowledgements}

Generative AI models were used to proofread and edit the text of this article.
The authors reviewed and take responsibility for all content, including the
system design, the implementation, and the experimental results and their
interpretation.

\bibliographystyle{ACM-Reference-Format}
\bibliography{references}

\end{document}